\documentclass[sigconf, nonacm]{acmart}

\AtBeginDocument{%
  \providecommand\BibTeX{{%
    \normalfont B\kern-0.5em{\scshape i\kern-0.25em b}\kern-0.8em\TeX}}}

\usepackage{array} 
\usepackage{pdflscape} 
\usepackage{longtable} 

\graphicspath{{./images/}}

\begin{document}

\title{Literature review on assistive technologies for people with Parkinson’s disease}



\author{Subek Acharya}
\authornote{Both authors contributed equally to this research.}
\email{subekacharya@uri.edu}
\author{Sansrit Paudel}
\orcid{0009-0004-4309-2870}
\authornotemark[1]
\email{sansrit@uri.edu}
\affiliation{%
  \institution{The University of Rhode Island}
  \city{Kingston, 02881}
  \state{Rhode Island}
  \country{USA}
}




\begin{abstract}

Parkinson's Disease (PD) is a neurodegenerative disorder that significantly impacts motor and non-motor functions. There is currently no treatment that slows or stops neurodegeneration in PD. In this context, assistive technologies (ATs) have emerged as vital tools to aid people with Parkinson's and significantly improve their quality of life. This review explores a broad spectrum of ATs, including wearable and cueing devices, exoskeletons, robotics, virtual reality, voice and video-assisted technologies, and emerging innovations such as artificial intelligence (AI), machine learning (ML), and the Internet of Things (IoT).

The review highlights ATs' significant role in addressing motor symptoms such as freezing of gait (FOG) and gait and posture disorders. However, it also identifies significant gaps in addressing non-motor symptoms such as sleep dysfunction and mental health. Similarly, the research identifies substantial potential in the further implementation of deep learning, AI, IOT technologies. Overall, this review highlights the transformative potential of AT in PD management while identifying gaps that future research should address to ensure personalized, accessible, and effective solutions. 
\end{abstract}


\keywords{Assistive Technologies, Parkinson’s Disease, Freezing of Gait (FOG), Virtual Reality, Wearable Devices, Exoskeletons, Cueing System, People with Parkinson's (PwP)}


\maketitle

\section{Introduction}
Parkinson's disease (PD) is a chronic, progressive neurodegenerative disorder that significantly impacts both motor and non-motor functions, making it the second most common neurodegenerative disease after Alzheimer’s. It affects approximately six million individuals globally, six hundred thousand to one million individuals in the US, with its prevalence expected to rise in the coming decades.\cite{dorsey2018} PD primarily results from the degeneration of dopamine-producing neurons in the substantia nigra, a region of the brain critical for facilitating movement. This degeneration disrupts motor pathways, leading to the disease's hallmark symptoms. 

\subsection{Symptoms of Parkinson's Disease}

PD manifests through a wide range of motor and non-motor symptoms, which progressively worsen over time.

\subsubsection{Motor Symptoms}
The primary motor symptoms of PD include:
\begin{itemize}
    \item \textbf{Bradykinesia:} 
    Bradykinesia is a typical and common symptom of PD. During bradykinesia, most patients show slowness of movement at onset, or experience reduced spontaneous movement, slow and clumsy, as well as reduced speed or amplitude of movement during rapid and repetitive movements\cite{dai2015} . Bradykinesia significantly impairs the quality of life of patients.
    
    \item \textbf{Tremor:}
    Tremor is often the earliest occurrence of Parkinson’s disease. It usually begins from one side of the upper limb, into the ipsilateral lower limb, and then spreads to the contralateral upper and lower limbs. It happens during rest, relaxes during voluntary movement, intensifies during nervous excitement, and disappears during sleeping. A rub-pill-like involuntary movement, with a frequency of 4–6Hz, typically characterizes the tremor\cite{elias2014}. Whereas tremor is not life-threatening, it can affect the patient’s ability to live a comfortable life by compromising various aspects such as dieting, writing, dressing, and self-care.
    
    \item \textbf{Postural Instability and Gait Disorder:}
    Postural Instability and Gait disorder are the main forms of postural balance impairment. Often, at the early stages of PD, the swing of the upper limbs on the affected side decreases, and the lower limbs drag. As PD progresses, freezing of gait (FOG), forward thrust gait, and panic gait develop, all of which make gait disorder the most disabling motor symptom \cite{mancini2019}\cite{tripoliti2013}.
    
    \item \textbf{Myotonia:}
    Myotonia is an impairment of muscle relaxation, similar to bending a soft lead tube, so-called “lead tube” rigidity. Myotonia often presents gear-like rigidity that affects many joints of the body and presents a special buckling posture. At the clinical level, the Unified Parkinson’s Disease Rating Scale (UPDRS) is often used to evaluate the neck, wrist, elbow, knee, ankle, or other joints of a patient\cite{goetz2008}. Another method includes quantitatively assessing myotonia based on the parameters extracted from the passive motion of the joints\cite{lee2002}.
\end{itemize}

\subsubsection{Non-Motor Symptoms}
In addition to motor impairments, PD is associated with various non-motor symptoms that significantly affect patients’ quality of life:
\begin{itemize}
    \item \textbf{Orthostatic hypotension:} Among Parkinson ’s-related cardiovascular abnormalities, orthostatic hypotension (OH) is the most common and one of the better-described conditions with an overall prevalence ranging from 10\% to 70\%. OH is multifactorial in origin, as it can be iatrogenic, but also an intrinsic feature of PD occurring early in the course of the disease \cite{bouhaddi2004}, as well as in the prodromal phase \cite{postuma2013}
    \item \textbf{Sleep Dysfunction:} Sleep dysfunction, such as onset and maintenance insomnia, are a common feature occurring in PD with wide range estimates regarding prevalence (20\% up to 80\%) reported in PD cross-sectional subjective assessments \cite{chahine2017}.
    \item \textbf{Impulse control disorder:} Impulse control disorders (ICD) are characterized by the inability to assert self-control in emotions and behaviors, leading to compulsive and/or impulsive actions that harm oneself or others. The pathways involved in ICD have been implicated in rapid eye movement sleep behavior disorder, constipation, and cognitive impairment \cite{augustine2021}.
    \item \textbf{Gastrointestinal:} Gastrointestinal dysfunction in PD affects the whole gastrointestinal tract and can be observed in each stage of the disease, from the prodromal to the advanced phases \cite{pfeiffer2018}. Especially constipation is a common feature with an overall prevalence ranging from 11 to 83\% and is considered one of the earliest symptoms of PD and a risk factor for the development of this condition \cite{adamscarr2016}.
    \item \textbf{Urinary:} Urinary symptoms are common in PD, as shown by studies where night-time urinary frequency (nycturia) was reported by 53\% of females and 63\% of male PwP, but also showing common urinary urgency and daytime frequency \cite{sakakibara2001}. Interestingly, several groups have separately reported that the presence of urinary symptoms at diagnosis is a significant biomarker for more rapid functional decline\cite{picillo2017}\cite{ayala2017}.
    \item \textbf{Depression:} Depression is the most common psychiatric complication in Parkinsons disease (PD) and affects 40–50\% of PD patients.Distinctive features of depression in PD include elevated levels of dysphoria, irritability, relative absence of guilt or feelings of failure and a low suicide rate despite a high frequency of suicidal ideation \cite{cummings1992}.

\end{itemize}

\subsection{Role of Assistive Technologies in Parkinson's Disease}
Assistive Technologies (AT) are defined as any item, piece of equipment, or product system, whether acquired commercially, modified, or customized, which is used to increase, maintain, or improve the functional capabilities of individuals with disabilities.\cite{assistive1998} 

Despite advancements in pharmacological and surgical treatments, current therapies for Parkinson's disease (PD) primarily manage symptoms but do not halt disease progression, creating an urgent need for complementary solutions. No current treatments can slow or stop the neurodegeneration associated with PD, highlighting the critical role of assistive technologies in improving the quality of life for individuals with the condition\cite{practical_neurology_2017}. 

These technologies offer the potential to aid mobility, rehabilitation, and daily tasks, addressing both motor and non-motor symptoms. Examples of AT include wearable and cueing devices, Robotics, Exoskeletons, Virtual Reality, Sensors, Internet Of Things and different emerging technologies. This review explores current technological solutions designed to aid early diagnosis of Parkinson's, alleviate symptoms, improve functional outcomes in PwP, and explore future possibilities of AT in the field of Parkinson's disease.

\section{Research Questions}

Based on the current scenario and the growing significance of assistive technologies in addressing the challenges faced by individuals with Parkinson’s disease (PD), the following research questions have been formulated to guide this study. These questions aim to explore the breadth and impact of assistive technologies in managing both motor and non-motor symptoms of PD, evaluating their effectiveness, and identifying future opportunities for innovation:

\begin{itemize} \item \textbf{What are the different types of assistive technologies available for assisting individuals with Parkinson’s disease?} \item \textbf{How effective are different categories of assistive technologies in managing Parkinson's disease motor and non-motor symptoms?} \item \textbf{What are the emerging trends and future directions in assistive technologies for Parkinson's disease rehabilitation?} \end{itemize}

These research questions aim to provide a comprehensive understanding of the current landscape of assistive technologies, their role in improving the quality of life for PD patients, and the opportunities for advancing these solutions to meet the unmet needs of this population.

\section{Methodology}

The methodology for this review focused on systematically identifying and analyzing academic literature on assistive technologies targeted toward alleviating the needs of individuals with Parkinson’s disease (PD). The process involved multiple steps, including keyword searches, inclusion, and exclusion criteria application, and systematic filtering to ensure the relevance and quality of the selected papers.

\subsection{Data Collection}

Relevant academic papers were identified using keyword-based searches on scholarly databases, particularly Google Scholar, Pub Med, ACM Digital Library, and Science Direct. The search queries were designed to include both general and specific terms to capture a comprehensive range of literature on assistive technologies for PD. The following keywords were used:

\begin{itemize} 
    \item \texttt{"Parkinson’s disease" AND "assistive technologies" AND "mobility"}
    \item \texttt{"Parkinson's disease" AND "gait rehabilitation"}
    \item \texttt{"Parkinson's disease" AND "cueing devices"}
    \item \texttt{"Parkinson's disease" AND "exoskeleton"}
    \item \texttt{"assistive devices" AND "Parkinson's disease" AND "quality of life"}
    \item \texttt{"Parkinson's disease" AND "wearable sensors"}
    \item \texttt{"Parkinson's disease" AND "robotics"}
    \item \texttt{"Parkinson's disease"AND "Virtual Reality"}
    \item \texttt{"Parkinson’s disease" AND "assistive technologies" AND "non-motor symptoms"}
    \item \texttt{"Parkinson’s disease" AND "assistive devices" AND "cognitive support"}
    \item \texttt{"Parkinson’s disease" AND "sleep dysfunction" AND "assistive technologies"}
    \item \texttt{"Parkinson’s disease" AND "Artificial Intelligence"}
\end{itemize}

\subsection{Inclusion and Exclusion Criteria}

To ensure the relevance and quality of selected papers, the following inclusion and exclusion criteria were applied:

\subsubsection{Inclusion Criteria} \begin{itemize} \item Papers published in or after 2015 were included to ensure the analysis of recent technologies. \item Full-length peer-reviewed articles focusing on Parkinson’s disease and assistive technologies. \item Studies specifically targeted to alleviate motor and non-motor symptoms of PD using Assistive technologies. \item Articles presenting detailed evaluations of assistive technologies. \end{itemize}

\subsubsection{Exclusion Criteria} \begin{itemize} \item Papers published before 2015. \item Short articles, book chapters, and opinion pieces were not included. \item Studies not specialized in addressing Parkinson’s disease. \item Studies that appeared multiple times due to being indexed in more than one database or published in different formats.

\end{itemize}

\subsection{Final Selection}

Following this rigorous screening process, the selected papers were categorized into various themes based on the technologies used. These papers provided a comprehensive overview of existing assistive technologies and their applications in the Parkinson's disease domain. 

This methodology ensured a focused and high-quality selection of studies for the literature review, providing insights into the state of assistive technologies for Parkinson’s disease and identifying gaps and opportunities for future research.

\section{Findings}

This section presents the findings from the literature review, categorizing assistive technologies designed to address motor and non-motor symptoms of Parkinson's Disease (PD). The technologies are classified based on their technologies and functionality.

\subsection{Wearable Devices and Cueing Systems}
Cueing can be defined as using external stimuli which provides temporal (related to time) or spatial (related to space) information to facilitate movement (gait) initiation and continuation. Wearable cueing devices have demonstrated significant efficacy in addressing freezing of gait (FoG) among Parkinson’s disease patients\cite{mazilu2015wearable}. The three primary types of cueing utilized are auditory, visual, and somatosensory cueing. While auditory cueing has been extensively researched over time, recent advancements have been made in visual and somatosensory cueing devices. All three cueing methods have proven effective in minimizing FoG episodes. However, usability concerns remain a common challenge across these devices\cite{zhao2016google}. Collectively, wearable cueing devices present promising solutions for improving mobility and quality of life, emphasizing the need for further development to enhance user experience and broader applicability.

Wearable devices, including smartwatches, smartphones, and specialized tools like the Parkinson’s KinetiGraph (PKG), have proven effective in monitoring and managing symptoms of Parkinson's disease. Smartwatches and smartphones can track both motor and non-motor symptoms, such as gait, tremor, and cognition, providing valuable insights through clinic-based and at-home monitoring over extended periods\cite{adams2023watchpd}. Similarly, the PKG effectively assessed sleep disturbances, including insomnia and restless legs syndrome, highlighting its utility for continuous symptom monitoring in both clinical and home environments\cite{klingelhoefer2016}. These technologies underscore the potential of wearable technologies in enhancing symptom management and patient care for Parkinson's disease.

Overall, advancements in wearable devices are seen in both motor and non-motor symptom management.

\begin{figure*}[t]
    \centering
    \includegraphics[width=\textwidth]{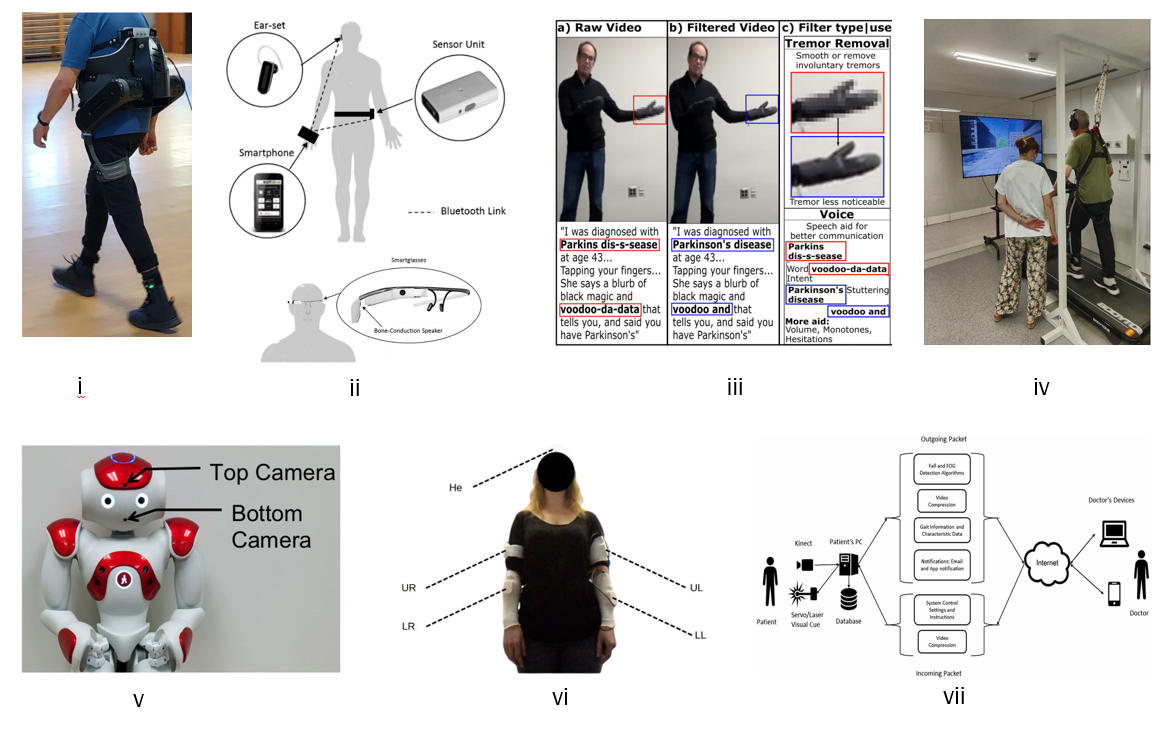}
    \caption{Overview of assistive technologies for Parkinson's disease: i. Active Pelvis Orthosis(Exoskeleton), ii. Auditory and visual cueing systems, iii. Voice and Video-Assisted Technology, iv. Gamified Virtual Reality Environment+treadmill set-up, v. Nao robot (SAR), vi. Participant wearing IMU Sensors, vii. Ambient-assisted living (AAL) environment with Microsoft Kinect v2 }
    \label{fig:devices}
\end{figure*}

\subsection{Exoskeletons}
Exoskeletons are wearable robotic devices designed to assist with physical movements, aiding gait rehabilitation by supporting movement initiation, balance, and walking stability \cite{otlet2023}\cite{romanato2022}. A study utilizing the Active Pelvis Orthosis reported notable improvements in hip range of motion, walking speed, and stride length, coupled with enhanced gait adaptability and variability, with benefits sustained for at least one month \cite{otlet2023}. Similarly, EksoGT® exoskeleton use has improved gait and postural stability, as evidenced by a randomized single-blind clinical trial involving 50 participants \cite{romanato2022}. Exoskeletons have been primarily used to alleviate motor symptoms, especially gait and postural stability.

\subsection{Robotics}
Robotic systems for PD fall into two main categories: 
\begin{itemize}
    \item \textbf{Socially Assistive Robots (SAR):} Socially Assistive Robots (SARs) have emerged as a promising area of research, primarily aimed at addressing the non-motor symptoms of Parkinson's disease. These robots focus on fulfilling the social and emotional needs of PwP by providing companionship, emotional support, and assistance with daily tasks. Research highlights that both patients and their families perceive SARs as valuable tools for enhancing emotional and social well-being, stressing the importance of user-centered design to align with their expectations and requirements \cite{kaplan2023}.
    
    Despite their potential, there are currently no SARs explicitly designed for PwP. Studies have identified significant challenges in developing SARs tailored to Parkinson’s patients, including understanding user-specific needs, ensuring safety and reliability, and designing intuitive and socially responsive behaviors \cite{wilson2020}.

The lack of SARs explicitly designed for Parkinson's patients and the technical challenges in their development present a critical area for future research and innovation.
\end{itemize}

\begin{itemize}
    \item \textbf{Robotic Rehabilitation Systems:} Robotic Rehabilitation System are used in motor function rehabilitation, helping Parkinson's patients improve gait and balance through structured, physical assistance during training sessions. Use of Tymo System and Walker view in improving gait and balance has seen promising results compare to traditional rehabilitation.  \cite{bevilacqua2020}
\end{itemize}

\subsection{Virtual Reality (VR)}
Virtual reality (VR) systems are primarily used in rehabilitation settings, focusing on gait abnormalities, static, and dynamic balance. VR has provided PwP a rehabilitation setting under dual-task conditions. Dual-tasking has been found to significantly improve mobility function \cite{bosch2024}\cite{tunur2020}. User satisfaction was observed, with participants favoring naturalistic environments and gamification elements for gait rehabilitation \cite{bosch2024}. Advanced VR systems, such as exergaming platforms with Kinect sensors, have shown significant improvements in both static and dynamic balance, outperforming conventional training methods \cite{brachman2021}. Furthermore, the integration of VR with motor imagery (MI) techniques, in combination with routine physical therapy, significantly enhanced motor function, balance, and activities of daily living (ADLs), with sustained improvements observed even at the 16-week follow-up \cite{kashif2022}. Overall, VR is seen as a promising field in rehabilitation for PwP.

\subsection{Voice and Video-Assisted Technologies}
Voice and Video-Assisted Technologies play a pivotal role in enhancing communication and managing symptoms for individuals with Parkinson's disease. Voice-assisted technologies (VAT), such as Alexa, Google Assistant, and Siri, have demonstrated potential in improving speech clarity, volume, and articulation, thereby offering therapeutic benefits to PwP and supporting daily tasks \cite{duffy2021}. Furthermore, wearable devices like EchoWear have proven effective in capturing high-quality speech data, facilitating the remote monitoring of speech exercises for PwP \cite{dubey2016}.

Similarly, PwP have shown positive attitudes toward the use of voice and video filters designed to reduce tremors and stabilize voice. These technologies have a significant positive impact on communication, reduce stigma, and substantially enhance the mental well-being of PwP \cite{haut2022}.

\subsection{Emerging Technologies}
Technologies like the Internet of Things (IoT), Artificial Intelligence (AI), and intent sensing are emerging in PD management. These technologies can be used in managing both motor and non-motor symptoms and also in diagnosis:
\begin{itemize}
    \item \textbf{AI and Intent Sensing:} Deep learning algorithms have shown significant success in predicting patient intent and symptom progression. For instance, intent-sensing probabilistic sensor networks (PSNs) have demonstrated high classification accuracy, underlining their potential for integration with assistive technologies like exoskeletons \cite{russell2023}. Similarly, decision tree classifiers have been proven to diagnose early symptoms of Parkinson's Disease with high accuracy, facilitating timely intervention and treatment \cite{yadav2023}.
   
    \item \textbf{IoT for Continuous Monitoring:} IoT-enabled systems facilitate real-time data collection and symptom monitoring, revolutionizing personalized healthcare for PD patients. Wearable technologies such as Kinesia, APDM, and PKG can be combined with IoT to enable real-time monitoring, diagnostics, and individualized treatment for Parkinson’s Disease \cite{pasluosta2015}. Moreover, IoT systems can help monitor and manage symptoms, improve medication adherence, and track emotional and social health. However, challenges such as usability and privacy concerns remain critical issues to address \cite{mcnaney2020}.
\end{itemize}

\subsection{Summary}

The findings revealed the transformative potential of assistive technologies in addressing both motor and non-motor symptoms as well as in early diagnosis of the disease. However, with some technologies, there are certain challenges of usability and practicality. The summary of studies on assistive technologies for Parkinson's Disease is further specified in Table 1.

\section{Discussion}

This research has successfully identified various assistive technologies for Parkinson's disease by addressing motor and non-motor symptoms.

\subsection{Types of Assistive Technologies Available for Parkinson’s Disease}

\begin{itemize}
    \item \textbf{Wearable Devices and Cueing Systems:} Primarily used for managing Freezing of Gait (FoG), monitoring motor symptoms, and addressing sleep disturbances.
    \item \textbf{Exoskeletons:} Focused on improving gait, mobility, and postural stability.
    \item \textbf{Robotic Systems:}
    \begin{itemize}
        \item \textit{Socially Assistive Robots (SARs):} Addressing emotional and social needs.
        \item \textit{Robotic Rehabilitation Systems:} Providing structured physical assistance for motor rehabilitation.
    \end{itemize}
    \item \textbf{Virtual Reality (VR) Systems:} Effective in improving gait, balance, and activities of daily living through immersive rehabilitation environments.
    \item \textbf{Voice and Video-Assisted Technologies:} Enhancing communication by reducing tremors, improving speech clarity, monitoring vocal stutters, and providing speech rehabilitation assistance.
    \item \textbf{Emerging Technologies:} IoT-enabled systems for real-time monitoring, AI for early detection, and modular sensing systems for intent sensing.
\end{itemize}

\begin{figure}[H] 
    \centering
    \includegraphics[width=0.45\textwidth]{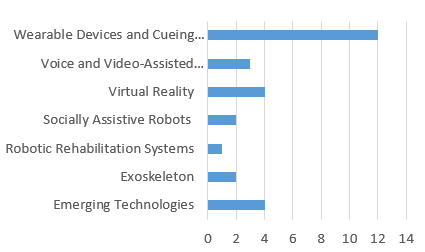} 
    \caption{Distribution of Research Papers Across Different Types of Assistive Technologies for Parkinson’s Disease}
    \label{fig:S2.png}
\end{figure}

\subsection{Effectiveness of Assistive Technologies in Managing Symptoms}

\begin{itemize}
   
    \item \textbf{Motor Symptoms:} ATs, particularly cueing devices, virtual reality, exoskeleton and robotic rehabilitation system have been successful in managing motor symptoms like Freezing of Gait (FOG), Gait and Posture Disorder. 
    \item \textbf{Non-Motor Symptoms:} Technologies like video and voice-assisted devices, IoT platforms, wearable, Socially Assistive Robots, AI   are emerging to address non-motor symptoms, such as speech impairments, sleep dysfunctions, early disease diagnosis, and emotional health. However, these areas are less researched compared to motor symptom management.
    
\end{itemize}

\begin{figure}[H] 
    \centering
    \includegraphics[width=0.45\textwidth]{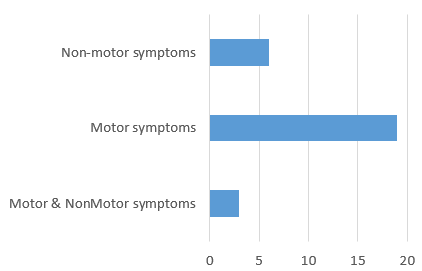} 
    \caption{Technologies addressing Motor vs Non-Motor Symptoms}
    \label{fig:Motor vs Non- Motor.png}
\end{figure}

\newgeometry{left=0.5in, right=0.5in, top=1in, bottom=1in} 
\begin{landscape}

\begin{longtable}{|p{2.5cm}|p{3cm}|p{3cm}|p{2.75cm}|p{4cm}|p{2.75cm}|p{2.75cm}|}
\hline
\textbf{Author(s) \& Year} & \textbf{Technology Category} & \textbf{Specific Device/Tool} & \textbf{Symptom Addressed} & \textbf{Key Findings/Outcomes} & \textbf{Study Design/Size} & \textbf{Application Setting} \\ \hline
\cite{mazilu2015wearable} Mazilu et al. 2015 & Wearable Devices and Cueing Systems & Auditory Cueing System (GaitAssist System) & Freezing of Gait (FoG) & Reduced FoG frequency and duration in 4 out of 5 participants; 97.1\% sensitivity for FoG detection with 0.5s lag time. Participants found the system wearable but noted setup challenges. & Proof of concept, n=5 (3-day study) \& subjective feedback from n=9 & Home-based gait training and daily life assistance \\ \hline
\cite{zhao2016google} Zhao et al. 2016 & Wearable Devices and Cueing Systems & Auditory Cueing System (Google Glass) & Freezing of Gait (FoG) & Significant immediate effect on mean frequency of EoD-FoG during 360° turning tasks (p < 0.05). Participants found the device easy to use and instructions clear but effect inconclusive due to small sample size. & Proof of concept, n=12 (lab study) & Home-based and lab-based gait assistance \\ \hline
\cite{zhao2016google} Zhao et al. 2016 & Wearable Devices and Cueing Systems & Visual Cueing System (Google Glass) & Freezing of Gait (FoG) & Visual cueing modes (optic flow and rhythmic flashing) showed no significant effect on FoG frequency/duration overall, but highlighted usability issues (e.g., distraction). Suggested placement of visuals centrally to improve usability. & Proof of concept, n=12 (lab study) & Home-based and lab-based gait assistance \\ \hline
\cite{tang2017rhythmic} Tang et al. 2017 & Wearable Devices and Cueing Systems & Visual Cueing System (Chest-worn Laser Device) & Freezing of Gait (FoG) & Rhythmic visual cueing significantly reduced On-FoG episodes (p = 0.001). Static visual cueing also reduced On-FoG but was less effective (p = 0.742). Generalizability limited by small sample size. & Experimental study, n=23 & Lab-based gait assistance \\ \hline
\cite{ahn2017smart} Ahn et al. 2017 & Wearable Devices and Cueing Systems & Visual Cueing System (Moverio BT-200 Smart Glasses) & Freezing of Gait (FoG) & FOGDOG algorithm achieved 97\% sensitivity and 88\% specificity for FoG detection with 1.1s lag. Adaptive visual cueing dynamically adjusted to gait speed and head orientation. Battery limitations noted. & Proof of concept, n=10 & Lab-based gait assistance \\ \hline
\cite{barthel2017laser} Barthel et al. 2017 & Wearable Devices and Cueing Systems & Visual Cueing System (Laser Shoes) & Freezing of Gait (FoG) & Significant immediate reduction in Off-FoG and On-FoG episodes and time spent in FoG (p < 0.01). 12/19 participants reported moderate to large improvement. High usability interest expressed. & Experimental study, n=19 & Lab-based gait assistance \\ \hline
\cite{mccandless2016cueing} McCandless et al. 2016 & Wearable Devices and Cueing Systems & Somatosensory Cueing System (BodyBeat Pulsing Metronome) & Freezing of Gait (FoG) & Positive effect on reducing Off-FoG episodes (mean Off-FoG percentage reduced from 81.58\% to 68.29\%). Generalizability limited due to task simplicity. & Experimental study, n=20 & Lab-based gait assistance \\ \hline

\cite{rosenthal2018sensory} Rosenthal et al. 2018 & Wearable Devices and Cueing Systems & Somatosensory Cueing System (cueStim Electrical Stimulator) & Freezing of Gait (FoG) & Statistically significant reduction in On-FoG episodes (58.28\% ± 33.89\%). Adjusted for sensory response without inducing motor response. & Home-based study, n=9 & Home-based gait assistance \\ \hline
\cite{goncalves2018parameters} Gonçalves et al. 2018 & Wearable Devices and Cueing Systems & Somatosensory Cueing System (Waistband Vibrotactile System) & Freezing of Gait (FoG) & Determined optimal vibration frequency (80–250 Hz, average sensitivity at 180 Hz) and minimum duration (>250 ms) for effective cueing. No FoG-specific testing performed. & Experimental study, n=30 (15 healthy, 15 PwP) & Lab-based vibration parameter testing \\ \hline
\cite{mancini2018cueing} Mancini et al. 2018 & Wearable Devices and Cueing Systems & Somatosensory Cueing System (VibroGait System) & Freezing of Gait (FoG) & Immediate reduction in time spent in Off-FoG (single task: 19 ± 18\%, dual task: 18 ± 15\%). Approximately 50\% of participants reported improvement in FoG. & Experimental study, n=25 & Lab-based gait assistance \\ \hline
\cite{adams2023watchpd} Adams et al. (2023) & Wearable Devices and Cueing Systems & Smartwatch (Apple Watch 4/5) and Smartphone (BrainBaseline™ App) & Early Parkinson's Disease Symptoms & Smartwatches and smartphones effectively monitored motor and non-motor features of early PD. Significant differences were detected in gait, tremor, finger tapping, speech, and cognition between PD patients and controls. & Multicenter observational study; n = 82 PD, 50 age-matched healthy controls & Clinic-based and at-home monitoring over 12 months. \\ \hline
\cite{klingelhoefer2016} Klingelhoefer et al. (2016) & Wearable Devices and Cueing Systems & Wearable Technology (Parkinson’s KinetiGraph (PKG)) & Night-time Sleep Disturbances in PD & PKG effectively assessed sleep disturbances such as insomnia and restless legs syndrome in PD patients. & Prospective comparative study; n = 63 PD patients (30 PD-EDS, 33 PD-NS) & Clinical and at-home monitoring over six consecutive days. \\ \hline
\cite{otlet2023} Otlet et al. 2023 & Exoskeleton & Active Pelvis Orthosis & Gait \& Postural Disorders & Improved hip range of motion, walking speed, and stride length. Enhanced gait adaptability and variability with sustained benefits after one month. & Experimental study, n=8 & Lab-based and overground gait training \\ \hline
\cite{romanato2022} Romanato et al. 2022 & Exoskeleton & EksoGT\textregistered{} Exoskeleton & Gait \& Postural Disorders & Improved gait and postural stability in Parkinson’s disease patients after exoskeleton-assisted gait training. & Randomized single-blind clinical trial, n=50 & Lab-based and clinical rehabilitation \\ \hline
\cite{kaplan2023} Kaplan et al. 2023 & Socially Assistive Robots & Not device-specific & Parkinson's management & Explored needs, attitudes, and concerns of individuals with Parkinson’s and their families regarding robotic technology. & Focus group study, n=15 (patients and family members) & Qualitative analysis of attitudes toward robotics \\ \hline
\cite{wilson2020} Wilson et al. 2020 & Socially Assistive Robots & Not device-specific & Social \& Mobility Support & Identified challenges in designing autonomous SARs for Parkinson’s, including understanding user needs, designing social behaviors, and ensuring safety/reliability. & Conceptual analysis & Assistive robot development \\ \hline
\cite{bevilacqua2020} Bevilacqua et al. 2020 & Robotic Rehabilitation Systems & Tymo System \& Walker View & Gait and Fall Risk in Parkinson’s & Protocol for RCT to compare traditional rehabilitation and robotic-based interventions. Aims to improve gait, balance, and reduce fall risks in older PD patients. & RCT, n=195 & Lab-based and clinical rehabilitation \\ \hline
\cite{bosch2024} Pere Bosch-Barceló et al. (2024) & Virtual Reality & Gamified Virtual Environment (Treadmill + Virtual Reality Environment) & Gait abnormalities in PD & Feasibility demonstrated with a SUS score of 74.82 ± 12.62 and a NATU Quest score of 4.49 ± 0.62. Participants favored naturalistic environments and appreciated gamification elements. & Proof of concept; 3 sessions with 4 PwPD \& 4 physiotherapists & Lab-based treadmill gait training \\ \hline
\cite{tunur2020} Tumay Tunur et al., 2020 & Virtual Reality & Google Glass with "Moving Through Glass" dance modules & Mobility, Dual-task Motor Function & Significant improvement in mobility under dual-task conditions; no significant changes in balance, quality of life, or mood; intervention was safe and well-tolerated. & Pilot study with 7 participants over 3 weeks & Home-based dance intervention using AR \\ \hline
\cite{brachman2021} Brachman et al., 2021 & Virtual Reality & Custom exergaming system with Kinect sensor and force platform & Static and Dynamic Balance & Exergaming significantly improved both static and dynamic balance. The exergaming group showed greater improvement in leaning rate (p=0.007) and dynamic balance (p=0.02) compared to conventional training. & Randomized controlled trial (n=24, 4-week program) & Clinical rehabilitation settings \\ \hline
\cite{kashif2022} Kashif et al., 2022 & Virtual Reality (VR) & Wii Fit System and MI techniques & Motor Function, Balance, ADLs & Combination of VR and MI with routine physical therapy (PT) significantly improved motor function (UPDRS-III), balance (BBS), balance confidence (ABCS), and ADLs (UPDRS-II) compared to PT alone. Improvements were retained at the 16-week follow-up. & Randomized controlled trial, n=44 (20 experimental, 21 control), 12-week intervention & Clinical rehabilitation setting \\ \hline
\cite{haut2022} Kurtis G. Haut et al., 2022 & Voice and Video-Assisted Technologies & Video and Voice filter prototype & Tremors (physical and vocal) & Filters to smooth physical tremors and adjust vocal stutters can significantly improve communication, reduce stigma, and positively impact mental health of PwP. & Surveys: 177 PwP, 107 general public; Interviews: 52 PwP, 3 health professionals & Virtual communication, video conferencing, social media \\ \hline
\cite{duffy2021} Orla Duffy et al., 2021 & Voice and Video-Assisted Technologies & Voice-Assisted Technology (VAT) (Amazon Alexa, Google Assistant, Apple Siri) & Speech and communication difficulties in PD & VATs can support communication, improve speech clarity and volume for some users, and increase independence. Privacy and usability concerns were minimal. & Survey: 290 respondents with PD; 166 VAT users; mixed-methods analysis, including VHI and qualitative content analysis. & Home settings, therapy support, daily tasks \\ \hline
\cite{dubey2016} Dubey et al., 2016 & Voice and Video-Assisted Technologies & Wearable Technology EchoWear Smartwatch & Speech and Voice Disorders in PD & EchoWear effectively recorded speech data comparable to high-quality laboratory equipment; shows promise for remote monitoring of speech exercises. & Pilot study; n=6 (3 PD patients, 3 healthy controls) & Remote monitoring of speech exercises prescribed by speech-language pathologists \\ \hline
\cite{russell2023} Joseph Russell et al., 2023 & Emerging Technologies & Modular intent sensing system (IMU sensors, microphone) & Intent prediction for PwP to assist daily tasks & Achieved 97.4\% accuracy in intent prediction within 0.5s and 99.9918\% over time; potential for assistive devices in improving task completion & 34 participants: 15 PwP, 19 controls; Activities: unlocking doors, buttoning cardigan, making toast & Controlled environment; Potential integration with assistive devices for PwP \\ \hline
\cite{pasluosta2015} Pasluosta et al., 2015 & Emerging Technologies & Wearable Technologies and IoT (Kinesia, APDM, PKG) & Motor symptoms (e.g., tremors, gait) \& Sleep Dysfunction \& Disease Diagnosis & Wearable technologies combined with IoT offer a lateralized healthcare platform enabling real-time monitoring, diagnostics, and individualized treatment for Parkinson’s Disease patients. & Review and discussion of existing systems and concepts & Home-based monitoring, real-time diagnostics, and treatment support \\ \hline
\cite{mcnaney2020} McNaney et al., 2020 & Emerging Technologies & Internet of Things (IoT) &  Mental Health \& Sleep Dysfunction \& Disease Diagnosis & Explored potential of IoT to monitor and manage symptoms, improve medication adherence, and track emotional/social health. Identified challenges of usability and privacy. & Multidisciplinary 2-day event with 23 professionals, 4 workshops with 13 PwP and caregivers (3 hours each) & Potential use in personal symptom tracking, medication adherence, and telemedicine applications. Focused on improving self-awareness, symptom management, and clinical decision-making. \\ \hline
\cite{yadav2023} Yadav, Singh, \& Pal (2023) & Emerging Technologies & Artificial Intelligence (Decision Tree Classifier) & Disease Diagnosis & Achieved 94.87\% accuracy in detecting PD using machine learning. Receiver Operating Characteristic (ROC) AUC was 98.7\%, validating strong performance. & Dataset with 23 features and 197 instances. 11 significant features identified through feature selection methods. & Clinical diagnosis and monitoring of Parkinson’s disease \\ \hline

\caption{Summary of studies on assistive technologies for Parkinson's Disease.}
\label{table:assistive_technologies}
\end{longtable}
\end{landscape}
\restoregeometry 

\twocolumn

\begin{figure}[H] 
    \centering
    \includegraphics[width=0.45\textwidth]{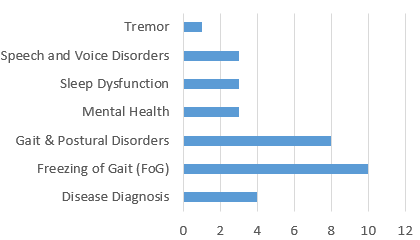} 
    \caption{Symptoms Addressed By Assistive Technologies}
    \label{fig:S2b.png}
\end{figure}

\subsection{Emerging Trends and Future Directions}

The use of AI and deep learning is a relatively new and promising area in Parkinson's Disease research. Although limited in scope, existing studies have demonstrated significant potential for future advancements and effective outcomes. Similarly, Socially Assistive Robots (SARs) represent another emerging field yet to be fully explored. However, current research highlights the challenges of developing tailored technologies specifically suited for PwP. Further studies are required to comprehensively evaluate the feasibility and effectiveness of SARs for PwP.

Additionally, IoT holds considerable promise as a prospect for mobile monitoring of Parkinson’s symptoms. Critical issues such as usability and security must be addressed to ensure its successful implementation and adoption.

\section{Limitations}

This research has certain limitations that needs to be acknowledged. Parkinson's disease presents a broad spectrum of symptoms, each of which could be analyzed in detail for the application of assistive technologies. However, this paper provides a general overview and does not explore the specific use of assistive technologies for each symptom in depth. Additionally, the study focuses on research published after 2015 to prioritize recent advancements and ensure the scope remained achievable within a limited timeframe. While this approach adds relevance, it potentially omits insights into the historical evolution of assistive technologies for Parkinson’s disease. Moreover, the paper relies on studies where assistive technologies were applied to Parkinson’s patients, with participant verification based on the descriptions provided in those studies. This reliance introduces a potential limitation, as it cannot be guaranteed that all participants were accurately diagnosed as people with Parkinson's, which might skew the findings.

Similarly, This review provides a limited exploration of assistive technologies targeting non-motor symptoms of Parkinson’s disease, as the inclusion criteria emphasized studies explicitly specialized in Parkinson's. This focus may have restricted a broader understanding of assistive technologies for non-motor symptom management. Lastly, many of the reviewed studies were conducted in laboratory or pilot settings, which may not fully represent the challenges and complexities of implementing these assistive technologies in real-world contexts.

\section{Conclusion}

This research has reviewed the available assistive devices for addressing the motor and non-motor symptoms of Parkinson’s Disease. The existing technologies include wearable devices and cueing systems, exoskeletons, robotics, virtual reality, voice and video-assisted technologies, and emerging technologies such as IoT, AI, and machine learning. Most of these devices focus on managing motor symptoms, such as freezing of gait (FoG), gait abnormalities, and postural instability. However, relatively few studies have focused on addressing non-motor symptoms, such as speech difficulties, emotional health, and sleep dysfunction.

Despite significant research progress, several challenges remain. Most studies have been conducted in laboratory settings, which do not accurately reflect real-world scenarios. Additionally, issues related to usability, accessibility, and privacy further limit the broader adoption of these technologies.

Emerging technologies like AI, IoT, and socially assistive robots (SARs) show promising potential for future research and development. However, their application in Parkinson’s management is still limited, and further research is needed to make these technologies more robust and effective in real-world settings.

Since Parkinson’s disease is a neurodegenerative disorder with no permanent cure, assistive devices play a crucial role in addressing the needs of PwP. By addressing current gaps and leveraging advancements in technology, future innovations can significantly enhance the quality of life for individuals living with Parkinson’s disease.


\bibliographystyle{ACM-Reference-Format}
\bibliography{references.bib}

\end{document}